\documentclass[twocolumn]{revtex4-2}

\usepackage[utf8]{inputenc}
\usepackage{bibunits}
\usepackage{amssymb,amsfonts,amsmath,mathrsfs,bbold}
\usepackage[normalem]{ulem}

\usepackage[usenames,dvipsnames]{xcolor}
\usepackage[colorlinks=true,citecolor=Blue,urlcolor=Blue,linkcolor=Blue]{hyperref}

\usepackage[all]{hypcap}
\usepackage{multirow}
\usepackage{graphicx}
\usepackage{dsfont}
\usepackage{braket}
\usepackage{chngcntr}

\newcommand{\aurelia}[1]{{\color{black} #1}}
\newcommand{\pablo}[1]{{\color{black} #1}}
\newcommand{\ruth}[1]{{\color{black} #1}}

\newcommand{\change}[1]{{\color{black} #1}}

\begin{document}

\title{Surmise for random matrices' level spacing distributions beyond nearest-neighbors}

\author{Ruth Shir}
\email{ruth.shir@uni.lu}
\affiliation{Department of Physics and Materials Science, University of Luxembourg, L-1511 Luxembourg, Luxembourg}
\author{Pablo Martinez-Azcona}
\affiliation{Department of Physics and Materials Science, University of Luxembourg, L-1511 Luxembourg, Luxembourg}
\author{Aur\'{e}lia Chenu}
\affiliation{Department of Physics and Materials Science, University of Luxembourg, L-1511 Luxembourg, Luxembourg}

\begin{abstract}
 Correlations between energy levels can help distinguish whether a many-body system is of integrable or chaotic nature. The study of short-range and long-range spectral correlations generally involves quantities which are very different, unless one uses the  $k$-th nearest neighbor ($k$NN) level spacing distributions. For nearest-neighbor (NN) spectral spacings, the distribution in random matrices is well captured by the Wigner surmise. This well-known approximation, derived exactly for a 2$\times$2 matrix, is simple and satisfactorily describes the NN spacings of larger matrices. There have been attempts in the literature to generalize Wigner's surmise to further away neighbors. 
 However, as we show, the current proposal in the literature fails to accurately capture numerical data.
Using the known variance of the distributions from random matrix theory, we propose a corrected surmise for the $k$NN spectral distributions. This surmise better characterizes spectral correlations while retaining the simplicity of Wigner's surmise. We test the predictions against numerical results and show that the corrected surmise is systematically more accurate at capturing data from random matrices. Using the XXZ spin chain with random on-site disorder, we illustrate how these results can be used as a refined probe of many-body quantum chaos for both short- and long-range spectral correlations.
\end{abstract}

\maketitle

\section{Introduction}
The most well-known and widely used probe of quantum chaos is the nearest-neighbor level spacing distribution of a system's spectrum \cite{Haake:2010fgh}. The \textit{Bohigas-Giannoni-Schmit} conjecture \cite{BGH_conjecture} states that quantum chaotic systems have\change{, in the bulk of their spectrum,} spectral statistics corresponding to random matrices. The spectral statistics belongs to a particular random matrix universality class, depending on certain basic symmetries of the physical system's Hamiltonian. In the bulk of the spectrum \cite{Haake:2010fgh}, the spectral statistics of a closed chaotic quantum system are expected to belong to one of the three universality classes of Hermitian random matrix theory (RMT), represented by the Gaussian Orthogonal, Unitary or Symplectic Ensembles (GOE, GUE or GSE, respectively). \change{We note that the edges of the spectrum may not exhibit universality and may depend on the nature of the underlying physical system \footnote{In some cases the edges of the spectrum may also exhibit universality, see e.g.~\cite{Buijsman_2022, Carro_2023}.}.} 
In contrast, the \textit{Berry-Tabor} conjecture \cite{BT_conjecture} states that spectral statistics \change{in the bulk of the spectrum} of closed integrable quantum systems have the same statistics as the Poisson point process on a line, i.e. behave as if their energies were sampled independently. \change{More specifically, the statement is that quantum systems with classically integrable counterparts will have such level spacing statistics. There are a few counter examples: an obvious one being the harmonic oscillator but also more complex examples exist, which show that certain classically integrable systems can show RMT statistics~\cite{Benet_2003, Hern_ndez_Quiroz_2010}. Generically however, the two conjectures hold even for quantum many-body systems which do not have a classical counterpart. To summarize,} the most important distinction between chaotic and integrable spectral statistics is the existence, or not, of \textit{repulsion} between energy levels \change{in the bulk of the spectrum}, respectively, which indicates correlations between the energies.  
The presence of level repulsion and the agreement with the random matrix description of the spectrum have been experimentally tested in heavy nuclei \cite{harvey_spacings_1958, garg_neutron_1964, camarda_statistical_1983, haq_fluctuation_1982}, chaotic systems with a semiclassical limit such as microwave billiards \cite{stockmann_quantum_1990, Stöckmann_1999} and, more recently, many-body setups \cite{Dag_many_body_2023, dong_measuring_2024}.

Wigner famously derived the distribution of spacings in a $2\times 2$ random matrix \cite{wigner_1951} and, though not exact, \change{cf.~Figure 2 in \cite{dietz1990taylor} and} \cite{Haake:2010fgh, Gaudin_1961, dietz1991level}, obtained a surmise that works surprisingly well for much larger random matrices \cite{rosenzweig_spacings_1958}. This is a simple function which has proven to be extremely useful in the study of quantum chaos. However, it probes only short-range correlations between eigenvalues. 
Longer-range spectral fluctuations are also very relevant to characterize quantum chaos. In particular, they are key to explain the full extent of the ramp of the Spectral Form Factor (SFF) \cite{Martinez_Shir_Chenu_SFF_2025}. Their study includes quantities such as the number variance \cite{Dyson_Mehta_1963}, the spectral rigidity \cite{Berry_spectral_rigidity_1985}, or the power spectrum \cite{corps_longrange_2021}, as well as the distributions of closest and farthest neighbors \cite{srivastava2018ordered}. 
Formal, exact results for the $k$th nearest neighbor ($k$NN) spacing distributions are discussed in \cite{mehta2004random, Mehta_Pandey_1983, Grimm_2004}, but these lack an explicit form. Other works \cite{abul-magd_wigner_1999, rao_higher-order_2020} suggest an approximate simple form, which, as we discuss below, does not accurately capture the numerical data. 

\change{In general, quantities such as the $k$NN distributions (including the nearest-neighbor level spacing distribution), the number variance, the spectral rigidity, and in some cases even the SFF \footnote{\change{To compare the SFF profile for a physical system with that of RMT it is necessary to unfold the spectrum first, cf.~\cite{Suntajs:2019lmb, Martinez_Shir_Chenu_SFF_2025}.}}, require a process of \textit{unfolding} to be applied to the spectrum. This process (described in Appendix \ref{App:unfolding}) removes the global density dependence and keeps only the fluctuations in the spectrum to be analysed. Once performed correctly, universality emerges and we are able to compare the spectral statistics of physical Hamiltonians with those of (the unfolded spectrum of) random matrices and the Poisson point process. For this reason, the \textit{spacing ratios} \cite{Oganesyan_2007, atas2013ratios, atas2013joint} are often used since they do not require unfolding to obtain universal results. However, the results for longer-range spacing ratios are still limited to small $k$ \cite{atas2013joint, tekur_higher-order_2018}. Thus, to understand long-range correlations among eigenvalues it is nonetheless useful to study the above-mentioned quantities, which require unfolding.}

In this work, we take a fresh look at the $k$th nearest neighbor spacing distributions, modeled with a Wigner-like surmise. Using the variance of these distributions, we correct the exponent of the surmise which leads to better agreement with the numerical data. We quantify \change{how well it fits} the data using the standard deviation goodness-of-fit between the distributions. 
We further test our new surmise on a physical many-body model, namely the XXZ spin chain with random on-site magnetic fields. The spectral statistics of this model changes as a function of the disorder strength of the random, local fields. More specifically, for a small \change{but non-zero} value of disorder strength, it corresponds to `chaotic' spectral statistics---matching GOE spectral statistics; while for a large value of disorder strength, it exhibits `integrable' spectral statistics---described by the Poisson distribution. This behavior was shown in several previous works, see e.g. \cite{serbyn2016spectral, bertrand2016anomalous, sierant2019level, sierant2020model, serbyn2017thouless, santos2018nonequilibrium, buijsman2019random, sierant2020thouless}, as well as in the framework of the study of many-body localization \cite{MBL_review}. Here we use the corrected surmise to study the $k$th nearest neighbor distributions for this spin chain in its different regimes and to obtain a more refined measure of quantum chaos. In particular, our new surmise provides a distribution for every $k$, which contains more information than scalar indicators such as the number variance, but which, unlike the form known in the previous literature, has the correct variance. This gives a valuable tool to study the breakdown of random matrix behavior for long-range spectral statistics in single- and many-body quantum chaotic systems.

This paper is organized as follows: in Section \ref{sec:kNN}, we describe and discuss the currently known results for the $k$th nearest neighbor distributions\aurelia{---which we also re-derive in App. \ref{App:surmise_derivation} in a formal way}. Based on the (known) variance of these distributions, we introduce a corrected surmise. In Section \ref{sec:RMT_numerics}, we test how well this surmise performs against numerical results for the three Gaussian ensembles of random matrix theory and compare this performance with that of the old surmise. In Section \ref{sec:XXZ_numerics}, we describe the XXZ spin chain Hamiltonian with random on-site disorder. We compute the $k$NN distributions in its different regimes as a function of the disorder strength and use it to show when this system deviates from RMT. Finally, we summarize and discuss our results in Section \ref{sec:summary}.

\section{The $k$NN spacing distributions}
\label{sec:kNN}

\begin{figure*}
    \centering    \includegraphics[width=0.3\linewidth]{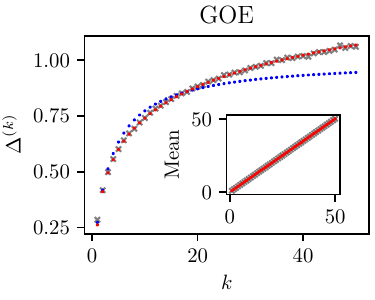} \hfill
    \includegraphics[width=0.3\linewidth]{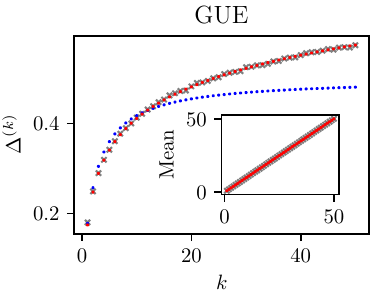} \hfill
    \includegraphics[width=0.3\linewidth]{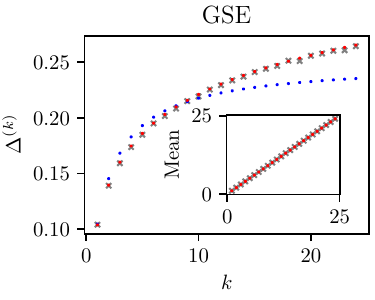}
    \caption{The variance and mean (inset) of the $k$th neighbor spacing distributions for the GOE, GUE, and GSE, for $k\geq 1$. The red dots show the analytical values using the new surmise, i.e. the exponent Eq.~\eqref{alpha_new} in the expression for the variance Eq.~\eqref{variance}. The blue dots show analytical values for the old surmise, i.e. using the exponent Eq.~\eqref{alpha_old} in Eq.~\eqref{variance}.
    The grey crosses show numerical data for 1000 realizations of random matrices of dimension $N=1000$ for the GOE and GUE and $N=2000$ for the GSE. 
    In all cases, the numerical mean for the unfolded spectrum satisfies $\langle s^{(k)} \rangle=k$. 
    }
    \label{fig:variance_and_mean}
\end{figure*}

\ruth{In this section we define and discuss the $k$th nearest neighbor spacing distributions, and provide a corrected Wigner-like surmise to describe them. }
\subsection{Wigner-like surmises}
The spectrum of a Hermitian matrix of size $N\times N$ consists of $N$ real eigenvalues, $E_1 \leq E_2 \leq \dots \leq E_N$.  The $k$th nearest neighbor spectral spacings are given by the set $\{s_i^{(k)}\}_{i=1}^{N-k}$, where \footnote{In what follows, we will omit the superscript $(k)$ for simplicity of notation, unless we want to explicitly highlight the role of $k$.} $s_i^{(k)}=E_{i+k}-E_i$. For $k=1$, these provide the distribution of nearest-neighbor (NN) spacings.  Hermitian random matrices can be categorized into three universality classes according to their fundamental symmetries. These classes markedly exhibit different behavior for their NN spectral spacings. The latter is given approximately by the Wigner surmise distribution \cite{Haake:2010fgh}
\begin{equation}
    \label{NN_WS}
    P^{(k=1)}(s) = C_\beta \, s^\beta \, e^{-A_\beta s^2},
\end{equation}
where $\beta=1,2,4$ for the Gaussian Orthogonal, Unitary and Symplectic ensembles, respectively. The coefficients $C_\beta$ and $A_\beta$ are set by normalizing the distribution to unity and normalizing the mean of the distribution to $\langle s\rangle = 1$.

While chaotic systems are expected to exhibit spectral statistics belonging to one of the universality classes of random matrix theory \cite{BGH_conjecture}, integrable systems are expected to have uncorrelated eigenvalues and to exhibit spectral statistics which coincide with those of the 1d Poisson point process \cite{BT_conjecture}. The $k$NN distributions for the Poisson point process are derived and given in App.~\ref{App:Pk_Pois}.
Here, we will be interested in finding a good approximation for the $k$th nearest-neighbor spacing distributions of the three universality classes of Hermitian random matrices. 

Wigner's surmise Eq.~\eqref{NN_WS} for NN spacings is found by considering the smallest possible random matrix with nearest-neighbor spacing, i.e. a $2\times 2$ matrix.  To find the $k$NN  spacing distribution, one may similarly attempt to consider a $(k+1)\times (k+1)$ sized matrix, as done in \cite{rao_higher-order_2020}. We perform such a computation in App. \ref{App:surmise_derivation} and find the following approximate behavior
\begin{equation}\label{Pk}
    P^{(k)}(s) \approx C_\alpha \, s^\alpha \, e^{-A_\alpha s^2},
\end{equation}
where the power $\alpha$ now depends on the spectral range $k$ as
\begin{equation}\label{alpha_old}
    \alpha = \frac{1}{2}k(k+1)\beta +k-1.
\end{equation}
Here as well, the coefficients $C_\alpha$ and $A_\alpha$ can be computed by setting the normalization of the distribution to 1, $\int_0^\infty P^{(k)}(s) \, ds = 1$ and setting the first moment of the distribution to $k$ by requiring $\langle s^{(k)}\rangle =\int_0^\infty s\,P^{(k)}(s) \, ds  = k$, where  $\langle \bullet \rangle \equiv \int_0^\infty P^{(k)}(s) \bullet ds$ denotes the average. 
This is expected for an unfolded spectrum \aurelia{(see App.~\ref{App:unfolding} for details on the unfolding procedures)} and can be verified numerically, see Figure \ref{fig:variance_and_mean} (insets). These conditions lead to 
\begin{equation}\label{A_and_C}
     A_{\alpha} = \left[\frac{\Gamma\left(\frac{\alpha}{2}+1\right)}{k\, \Gamma\left( \frac{\alpha+1}{2} \right)}\right]^2, \quad C_{\alpha} = \frac{2 A_\alpha^{\frac{\alpha+1}{2}}}{\Gamma\left(\frac{\alpha+1}{2} \right)} .
\end{equation}  

The evaluation of Gamma functions for large values of the argument is numerically unstable. For this reason, it will prove useful to introduce approximate expressions for $A_\alpha$ and $C_\alpha$ given in Eq.~\eqref{A_and_C} for large $\alpha$ as
\begin{align} 
    A_\alpha &= \frac{1}{k^2}\left(\frac{\alpha}{2 } + \frac{1}{4 } + \frac{1}{16  \alpha} \right) +O(\alpha^{-2}), \label{A_approx} \\
    C_\alpha &=\!\! \sqrt{\frac{\alpha}{\pi}}\frac{1}{k^{1+\alpha}}e^{\frac{1}{4}+\frac{\alpha}{2}+O(\alpha^{-2})}\!\left[1\!+\!\frac{1}{12\alpha}\!+\!O(\alpha^{-2}) \right]\nonumber\\
    &\approx \frac{1+12 \alpha}{12 \sqrt{\pi \alpha}\,k^{1+\alpha}}e^{\frac{1}{4}+\frac{\alpha}{2}} ~. \label{C_approx}
\end{align}
These approximations work very well, even for not very large $k$, since from Eq.~\eqref{alpha_old}, $\alpha$ is of $O(k^2)$. The corrected exponent that we define in Eq.~\eqref{alpha_new} behaves as $O(k^2/\ln k)$, for which these approximations still work accurately. For the simulation, we explicitly state whether we use the approximated or exact expressions.

\begin{figure*}[t]
    \centering
    \includegraphics[width=0.33\linewidth]{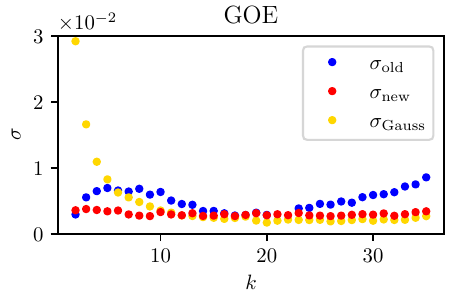}\hfill
    \includegraphics[width=0.33\linewidth]{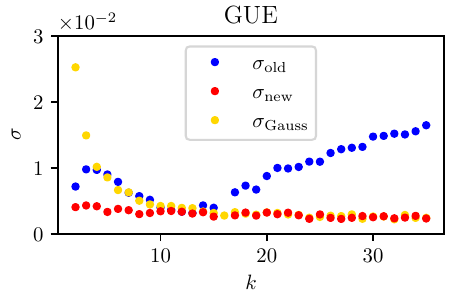}\hfill
    \includegraphics[width=0.33\linewidth]{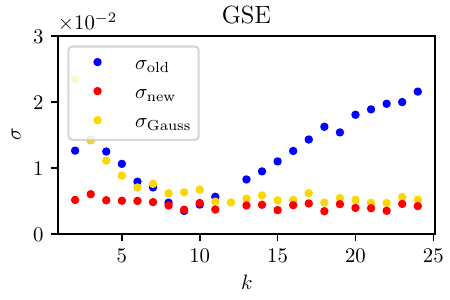}
    \caption{The standard deviation (goodness of fit), $\sigma$, defined in Eq.~\eqref{SD}, for $k\geq 2$ computed for the old surmise (blue), the new surmise (red) and the Gaussian surmise (gold) against numerical data of 1000 realizations of random matrices of dimension $N=1000$ for the GOE and GUE and $N=2000$ for the GSE. The new surmise Eq.~\eqref{Pk_new} systematically better captures the numerical data in the three ensembles.}
    \label{fig:SD}
\end{figure*}

The distribution Eq.~\eqref{Pk} with $\alpha$ given by Eq.~\eqref{alpha_old} was introduced in several works, see e.g. \cite{abul-magd_wigner_1999, rao_higher-order_2020, engel_higher-order_1998,  tekur_higher-order_2018, sakhr_wigner_2006, forrester_random_2009} \aurelia{and App.~\ref{App:surmise_derivation} for a discussion}, and argued for in various ways; we will refer to it as the \textit{old surmise}. It does recover the NN Wigner surmise by setting $k=1$ in the expression Eq.~\eqref{alpha_old} for $\alpha$ and using the latter to get the constants in Eq.~\eqref{A_and_C}.
 However, it does not work very well for neither small nor large $k$, as we discuss below. In particular, the variance
\begin{equation}\label{variance}
    \Delta^{(k)} =\langle {s^{(k)}}^2\rangle-\langle s^{(k)} \rangle^2= \frac{\alpha+1}{2A_\alpha} - k^2,
\end{equation}
with $\alpha$ given by Eq.~\eqref{alpha_old} does not follow the variance found numerically, as can be seen in Figure \ref{fig:variance_and_mean}. Indeed, the predictions from Eq.~\eqref{alpha_old} underestimate the $k$NN variance obtained numerically for random matrices (grey crosses)\pablo{, especially for large $k$, while for small $k$ they slightly overestimate the variance}. More specifically, we \pablo{will show that this distribution} Eq.~\eqref{Pk} \pablo{does not accurately model the $k$NN distribution for random matrices}---see Eq.~\eqref{SD} for the definition \pablo{of the goodness-of-fit indicator} and Figures \ref{fig:SD} and \ref{fig:Pk_plots} for the results.

Below, we discuss the exact variance and use it to correct the exponent $\alpha$. This yields a new surmise that better captures the distributions for the $k$NN spacings, as we verify against numerical data.

\subsection{The variance of the $k$NN spacing distributions}
While there seems to be no satisfactory (i.e. accurate \textit{and} simple) expression for the $k$NN spacing distributions of random matrices in the literature, the variance of these distributions is known \cite{french1978statistical, RevModPhys.53.385} and reads 
\begin{equation}\label{Delta_RMT}
    \Delta^{(k)}_\beta = 
   \frac{2}{\pi^2 \beta} \ln k +c_\beta .
\end{equation}
The coefficient $c_\beta$ can be thought of as the \textit{`boundary condition'} for $\Delta^{(k)}_\beta$ at $k=1$. 
\ruth{Following \cite{french1978statistical}, who proposed using the variance of Wigner's surmise Eq.~\eqref{NN_WS} as the boundary condition for GOE, we use the variance of Wigner's surmise also for GUE and GSE to compute the constant for each ensemble:}  
$c_1=4/\pi-1$, $c_2= 3\pi/8 - 1$ and $c_4 = 45\pi/128 - 1$.

It was observed \cite{french1978statistical} that the variance of the $k$NN distribution is related (to good precision) to the \textit{number variance} $\Sigma^2$ through 
\begin{equation}
    \label{eq_Sigma_Delta}
   \Delta^{(k)} =  \Sigma^2(k)-1/6~, 
\end{equation}
the number variance being defined as \cite{Dyson_Mehta_1963, Guhr:1997ve}
\begin{eqnarray}
    \Sigma^2(L) = \overline{ \eta^2(\xi_s, L)} - \overline{ \eta(\xi_s, L)}^2,
\end{eqnarray}
where $\eta(\xi_s, L)$ counts the number of levels in the interval $[\xi_s, \xi_s+L]$, where $L$ is a continuous parameter. The average, denoted by the overline, is taken over the starting points $\xi_s$. In practice, one also averages over realizations of the model. This relationship Eq.~\eqref{eq_Sigma_Delta} is not fully understood in the literature but was found to hold for various ensembles, including the GOE, GUE and GSE \footnote{Note that from this relationship, the constant $c_\beta$ can be read off as $c_1= \frac{2}{\pi^2}[\ln 2\pi +\gamma +1 -\pi^2/8] -1/6$ for GOE, $c_2=\frac{1}{\pi^2}[\ln 2\pi +\gamma +1 ] -1/6$ for GUE and $c_4= \frac{1}{2\pi^2}[\ln 4\pi +\gamma +1 +\pi^2/8] -1/6$ for GSE, where $\gamma\approx 0.57721$ is Euler's constant. \ruth{We have chosen, however, to use as the boundary condition the variance at $k=1$ as given by Wigner's surmise, since this choice was reported in \cite{french1978statistical} to work more accurately in the GOE case.}}.

Let us stress the difference between $\Delta^{(k)}$ and $\Sigma^2(L)$: the first measures the fluctuations in the \textit{length} of an interval containing $k+1$ levels (one at each end of the interval and $k-1$ within the interval), while the second measures the fluctuations in the \textit{number of levels} contained in an interval of length $L$. Since the spectrum is unfolded, the average energy difference between two consecutive levels is set at $\langle s^{(1)}\rangle = 1$, which sets these two quantities on the same scale. \change{We also note that the numerical computation of the $k$NN distributions and thus of $\Delta^{(k)}$ is more straightforward than that of the number variance, $\Sigma^2(L)$. While the $k$NN distributions require a simple difference operation between eigenvalues for every integer $1\leq k\leq k_{\text{max}}$, the computation of the number variance requires counting the number of eigenvalues that lie in every spectral window of size $L$ for every $0\leq L \leq L_{\text{max}}$. In Appendix~\ref{App:NV_vs_variance} we discuss the computational cost of computing numerically the number variance vs the variance of the $k$NN distributions, showing that computing the number variance is more costly. }

For comparison, the variance of the $k$NN distributions for the 1d Poisson point process is given by
\begin{eqnarray}\label{Delta_Poisson}
    \Delta^{(k)}_\text{Poisson} = k ~.
\end{eqnarray}
It is interesting to note that, for Poisson, in the relationship between the number variance and the variance of $P^{(k)}$, the $1/6$ disappears, and
\begin{eqnarray} \label{eq_Sigma_Delta_Pois}
    \Delta^{(k)}_\text{Poisson}=\Sigma^2_\text{Poisson}(k)~.
\end{eqnarray}
Later in Section \ref{sec:XXZ_numerics}, we will check the relationships Eq.~\eqref{eq_Sigma_Delta} and Eq.~\eqref{eq_Sigma_Delta_Pois} for a physical model with a transition between chaos and integrability.

Thus, for an unfolded spectrum, the mean of each $k$NN distribution always equals $k$. In turn, the variance of the distribution captures the different behaviors between Poisson and RMT eigenvalues' statistics---Eq.~\eqref{Delta_Poisson} shows a linear dependence $\sim k$ for Poisson, while Eq.~\eqref{Delta_RMT} shows it grows more slowly as $\sim \ln k$ for RMT. This is a manifestation of spectral rigidity in random matrices. Indeed, the distance between an eigenvalue and its $k$th NN is not as spread out as in Poisson, where eigenvalues are not correlated and the mean and the variance are equal to each other. 
Below, we use the knowledge of the variance to get a better surmise for the $k$NN spacing distributions.

\subsection{The corrected $k$NN spacing distribution}
The expression Eq.~\eqref{Pk} with $A_\alpha$ and $C_\alpha$ given by Eq.~\eqref{A_and_C} for the $k$NN spacing distributions, has  $\alpha$ as the only free parameter. Indeed, its variance Eq.~\eqref{variance} depends only on $\alpha$. We propose to correct $\alpha$ using the exact expression for the variance, Eq.~\eqref{Delta_RMT}. The corrected value, which we denote with a tilde as $\tilde{\alpha}$, is thus found by setting
\begin{equation} \label{Eq_for_talpha}
    \frac{\tilde{\alpha}+1}{2A_{\tilde{\alpha}}} - k^2 = \frac{2}{\pi^2 \beta} \ln k +c_\beta ~.
\end{equation}

Using the approximated expression Eq.~\eqref{A_approx} up to and including $O(\alpha^{-1})$, an explicit expression for the corrected exponent follows as 
\begin{align}
    \label{alpha_new_long}
    \!\tilde{\alpha}=& \frac{1}{{4 (\pi ^2 \beta  c_\beta+2 \ln k)}}\Big[{\pi ^2 \beta  k^2 -(\pi^2\beta c_\beta+2 \ln k)} \\
    &+\pi^2\beta k^2 \sqrt{1-\frac{\pi^2\beta c_\beta+2 \ln k}{\pi^2\beta k^2}\left(4- \frac{\pi^2\beta c_\beta+2 \ln k}{\pi^2\beta k^2}\right)}\Big].\nonumber
\end{align}
\change{In the large $k$ limit, the ratio $\frac{\pi^2\beta c_\beta+2 \ln k}{\pi^2\beta k^2}$ is small, of $O(k^{-2}\ln k)$. We can thus expand the square-root as $\approx1-
\frac{1}{2}\left[\frac{\pi^2\beta c_\beta+2 \ln k}{\pi^2\beta k^2}\left(4- \frac{\pi^2\beta c_\beta+2 \ln k}{\pi^2\beta k^2}\right)\right] \approx 1-2\frac{\pi^2\beta c_\beta+2 \ln k}{\pi^2\beta k^2}$. Using this approximation in Eq.~\eqref{alpha_new_long} we find that at large $k$, $\tilde \alpha$ can be approximated by:}
\begin{eqnarray}
    \tilde{\alpha}   = \frac{\pi ^2 \beta k^2}{2 \left(\pi ^2 \beta c_\beta+2\ln k\right)}-\frac{3}{4}+\change{O(k^{-2}\ln k)},\label{alpha_new}
\end{eqnarray}
with $c_\beta$ given under Eq.~\eqref{Delta_RMT}. 
Note that the $k^2$ behavior seen in Eq.~\eqref{alpha_old} now has a $\ln k$ correction in Eq.~\eqref{alpha_new}. Also, unlike the power of the old surmise, $\alpha$ in Eq.~\eqref{alpha_old}, the corrected exponent $\tilde \alpha$ is not necessarily an integer.
For NN spacings ($k=1$), this corrected power reads $\tilde \alpha = \frac{1}{2 c_\beta}-\frac{3}{4}$. It numerically reduces to $\tilde \alpha \approx 1.08, \, 2.06, \, 4.04$ for GOE, GUE and GSE, respectively, which are quite close to the values of $\beta$ for each of the corresponding ensembles. Therefore, even for NN where the expansions are not guaranteed to work, the numerical value of $\tilde \alpha$ is still reasonably close to the expected value. \change{In Appendix \ref{App:Pk1_WS} we study further the case of $k=1$ for the GUE.  Using the variance of the exact NN distribution, Eq.~\eqref{Ps_exact_GUE}, we solve Eq.~\eqref{variance} numerically to find a corrected power $\tilde \beta$, Eq.~\eqref{beta_new}. We show that using this corrected power in Eq.~\eqref{NN_WS} performs better than Wigner's surmise (with $\beta=2$) by comparing with numerical data and with the exact analytical result, see Figure~\ref{fig:Pk1}.}

The \textit{new surmise} is thus given by Eq.~\eqref{Pk} with $\alpha$ given by Eq.~\eqref{alpha_new} rather than Eq.~\eqref{alpha_old}, namely 
\begin{equation}\label{Pk_new}
    \tilde{P}^{(k)}(s) = C_{\tilde{\alpha}} \, s^{\tilde{\alpha}} \, \exp(-A_{\tilde{\alpha}}s^2)~.
\end{equation}
It is legitimate to ask how well this corrected surmise performs. To start with, Figure \ref{fig:variance_and_mean} compares the variance Eq.~\eqref{variance} as computed using the new Eq.~\eqref{alpha_new} or old Eq.~\eqref{alpha_old} exponent \pablo{with the variance obtained numerically for random matrices}. The constant $A_{\alpha}$ was computed using the exact expression for $k<10$ in both cases. We see that the variance of the new surmise (red dots) now matches numerical data for the variance as a function of $k$ for the three Gaussian ensembles. 
In the next Section, we go beyond the variance and quantify the performance of the new surmise by measuring the distance to the numerically generated distributions from RMT. We will see that, although it was derived using the large $\alpha$ approximation of $A_\alpha$, the surmise Eq.~\eqref{Pk_new} works well already for $k\geq 2$.

\subsection{Gaussian surmise for large k}
For large $k$, the $k$NN distributions are known to become Gaussian  \cite{RevModPhys.53.385}. \change{This can be seen for Eq.~\eqref{Pk_new} (or in the same way for Eq.~\eqref{Pk}) from Eq.~\eqref{variance} by noting that, at large $k$, the inflection points are located symmetrically around the mean at a distance given by the asymptotic value of the standard deviation, $\sqrt{\Delta^{(k)}}$. In particular the inflection points are given by $k\pm \frac{k}{\sqrt{2\tilde\alpha}}$. Here, $k$ is the mean, and $\frac{k}{\sqrt{2\tilde\alpha}}$ is $\sqrt{\Delta^{(k)}}$ at large $\tilde\alpha$. In addition, it can be checked that the asymptotic values of the Skewness and excess Kurtosis are given by $\mathrm{Skew} \to \frac{1}{\sqrt{2 \tilde\alpha}}$ and $\mathrm{Kurt} \to  \frac{3}{4 \tilde\alpha^2}$ which approach zero for large $\tilde\alpha$, showing that the distribution is becoming more and more Gaussian, see App.~\ref{App:Skew_Kurt}. The numerical data confirms that} the higher moments, namely the Skewness and excess Kurtosis, tend to zero for large $k$, see Figure \ref{fig:skew_and_kurt}. 

This behavior is captured by
a Gaussian centered at $k$ with a variance $\Delta^{(k)}$  as given in Eq.~\eqref{Delta_RMT}, i.e.
\begin{equation}\label{Pk_Gauss}
    \tilde{P}^{(k)}_\textrm{Gauss}(s) = \frac{1}{\sqrt{2 \pi \, \Delta^{(k)}}} \exp\left[- \frac{(s-k)^2}{2 \Delta^{(k)}} \right]~.
\end{equation}
\change{We now present some arguments in favor of Eq.~\eqref{Pk_Gauss}, following Appendix N of \cite{RevModPhys.53.385}. Each $k$NN spacing is a sum of $k$ NN spacings, namely $s_i^{(k)}=s_i^{(1)}+s_{i+1}^{(1)}+\dots+s_{i+k-1}^{(1)}$. Since the $s_i^{(1)}$ are identically distributed, in the absence of correlations between them, the central limit theorem would imply that the $k$NN distributions are Gaussian at large $k$. There are, however, correlations between the NN spacings. Yet, for NN spacings which are sufficiently far apart, the correlations fall off as $C(s_i^{(1)}, s_j^{(1)})\sim |i-j|^{-2}$ \cite{RevModPhys.53.385}, which means they can be considered \textit{uncorrelated} for $|i-j|\geq m$, thus constituting an $m$-\textit{dependent stationary sequence} \cite{Diananda_Bartlett_1953}. The $k$NN spacings, being a sum of such an $m$-dependent sequence, are predicted to have a Gaussian distribution for large $k$ according to the \textit{Diananda theorem} \cite{Diananda_Bartlett_1953}.}
We test also the Gaussian surmise in the next Section and find good agreement with numerical results for large enough $k$.

\begin{figure*}
    \centering
    \includegraphics[width=1\linewidth]{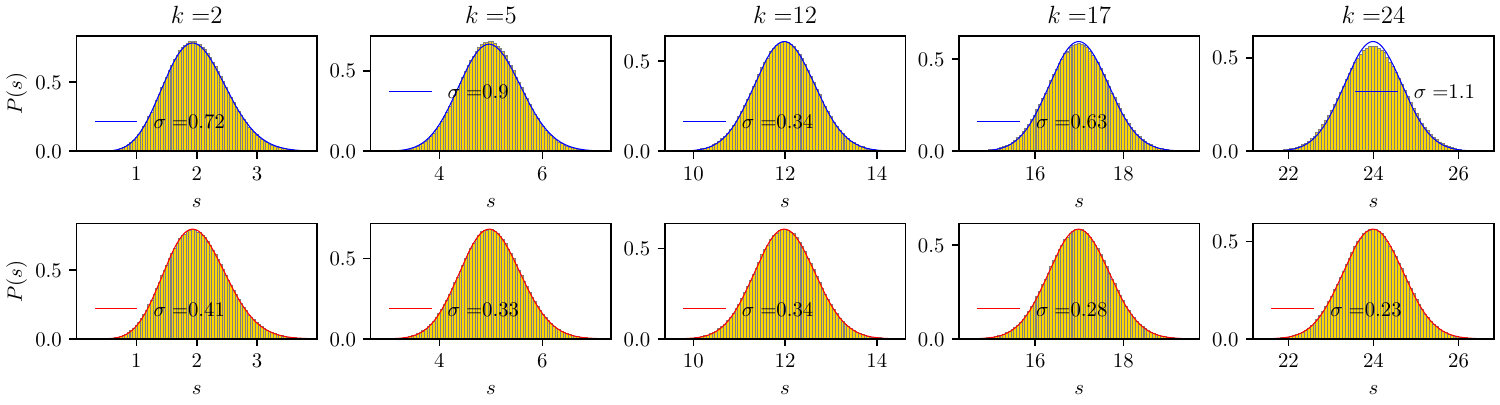}
    \caption{The $k$NN distributions for a few values of $k$, for the GUE. The histograms show the numerical data for 1000 realizations of $N=1000$ matrices taken from the GUE. The top row shows the old surmise (blue curves), while the bottom row shows the new surmise (red curves). \ruth{The standard deviation goodness-of-fit, $\sigma$, between the analytical distributions and the numerical data is shown in units of $10^{-2}$. See Appendix \ref{App:GOE_GSE} for similar data for the GOE and GSE.}}
    \label{fig:Pk_plots}
\end{figure*}

\section{Comparison with numerical results}
\label{sec:RMT_numerics}

In this Section, we compare the analytical surmise for the $k$NN distributions against numerical results. The analytical functions are (i) the old surmise Eq.~\eqref{Pk} with $\alpha$ given by Eq.~\eqref{alpha_old}, (ii) the corrected surmise given by Eq.~\eqref{Pk_new}, and (iii) the Gaussian surmise Eq.~\eqref{Pk_Gauss}, expected at large $k$.  The numerical results were generated from 1000 realizations of matrices of size $N=1000$ taken from the GOE and the GUE, and of size $N=2000$ for the GSE. The unfolding procedure we use is described in App.~\ref{App:unfolding}.

We test how well two distributions, $p_1(s)$ and $p_2(s)$, agree with each other
using the \textit{standard deviation} goodness-of-fit \cite{akemann_spacing_2022, akemann_transitions_2025}
\begin{equation} \label{SD}
    \sigma = \sqrt{\frac{1}{n}\sum_{i=1}^n [p_1(s_i)-p_2(s_i)]^2}.
\end{equation}
Here, $n$ in the number of bins and $s_i$ corresponds to the middle of the $i$th bin. Note that very small values in the tails of the distributions can result in a smaller value for $\sigma$ without reflecting a better fit.  To avoid this effect, we test the performance of the fit restricting ourselves to 3 standard deviations from the mean of the distribution.

Figure \ref{fig:SD} shows, for the GOE, GUE and GSE, the standard deviation Eq.~\eqref{SD} between the numerical values and either the old surmise, Eq.~\eqref{Pk} with $\alpha$ given by Eq.~\eqref{alpha_old}, the new surmise Eq.~\eqref{Pk_new}, or the Gaussian surmise Eq.~\eqref{Pk_Gauss}. For the old \change{and new} surmise\change{s}, we use the exact expressions for $A_\alpha$ and $C_\alpha$ for $k<10$ and the approximated ones for $k\geq 10$, for all three ensembles.

Note that the absolute values of $\sigma$ are not physically relevant because they may be affected by factors such as the values of the distributions and the number of bins. However, the value of $\sigma$ for $k=2$ in GOE, which is known to be \aurelia{exactly} equivalent to the nearest neighbor Wigner surmise in GSE \cite{Mehta1963}, may be taken as a benchmark value indicative of a `good' fit. We observe that the corrected surmise consistently performs as well or better, i.e. has a smaller value of $\sigma$ which is close to the benchmark\aurelia{---$\sigma=0.3 \times 10^{-2}$ for $\beta = 1$ and $k=2$---}than the old surmise. 
This means that the corrected surmise provides a better characterization of spectral correlations for $k\geq 2$ than the old surmise. 

It is noteworthy that the old surmise gives a good fit for a few intermediate values of $k$---as good as the corrected surmise---and otherwise provides a very poor fit for long-range spectral correlations. This behavior may be explained looking at Fig. \ref{fig:variance_and_mean}, which shows that the variance of the old surmise and the $k$NN distribution coincide at some particular $k$---numerically estimated to be $k \approx 18, \, 12, \, 8$ respectively for $\beta = 1, \, 2, \,  4$. These values also correspond to the points at which the old surmise becomes as good as the corrected one, cf. Fig. \ref{fig:SD}.  
We also see in Fig.~\ref{fig:SD} that the Gaussian surmise performs similarly well---slightly better for GOE, equally for GUE and slightly worse for GSE--- as the corrected surmise for long-range  $k \gtrsim 10$ neighbors.

\begin{figure*}
    \centering
    \includegraphics[width=1\linewidth]{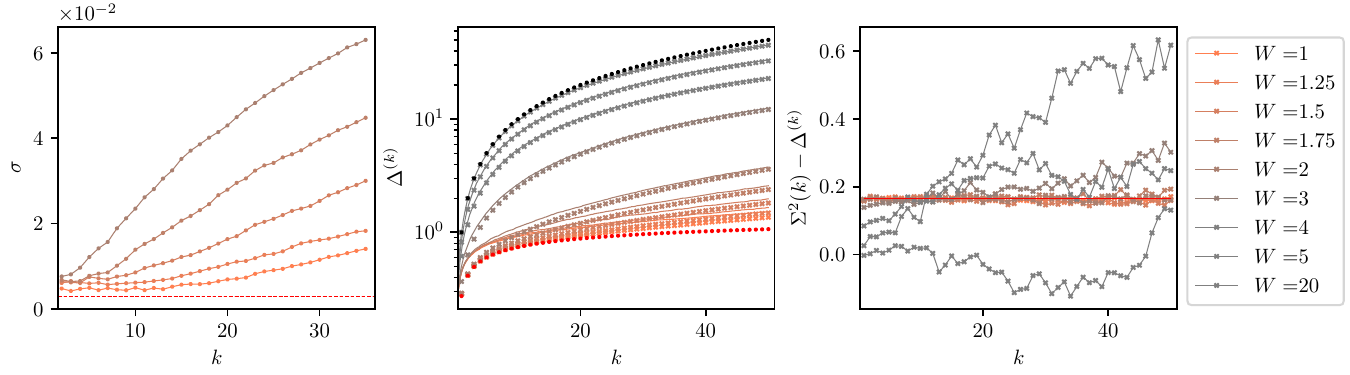}
    \caption{Results for the XXZ spin chain with random on-site magnetic fields. (Left) The goodness-of-fit, $\sigma$, for $W=1, 1.25, 1.5, 1.75$ and $2$, compared with the $k$NN GOE distributions \ruth{for $2\leq k \leq 35$. The dashed red line at $\sigma=0.3\times 10^{-2}$ is the benchmark value discussed in the main text; the connecting lines are shown as a guide to the eye.} (Middle) The variance of the $k$NN distributions for the disorder strengths $W$ listed in the legend \ruth{(log-scale on the $y$-axis)}.  For reference we added the variances Eq.~\eqref{Delta_RMT} for the GOE (red dots) and the variances  Eq.~\eqref{Delta_Poisson} for the 1d Poisson point process (black dots). The number variance is plotted as a continuous line. (Right) The difference between the Number Variance and the variances of the $k$NN distributions (crosses; the connecting lines are shown as a guide to the eye). The value of $1/6$ is plotted as a horizontal red line. 
    The full distribution histograms are shown for a few values of $k$ in Figure \ref{fig:XXZ_kNN}.}
    \label{fig:XXZ_variance}
\end{figure*}

To illustrate the different distributions, Fig.~\ref{fig:Pk_plots} displays the numerical histograms for $P^{(k)}(s)$ for GUE (results for the GOE and GSE can be found in App.~\ref{App:GOE_GSE}) for a few values of $k$, together with the old surmise (top) and corrected surmise (bottom). The goodness of fit $\sigma$ for each analytical distribution is also indicated. Again, we observe that the corrected surmise consistently provides a better fit of the numerical values. For small $k=4,\,6$, the old surmise predicts a slightly broader distribution than the numerical one, leading to a smaller value than expected in the peak of the distribution. For $k=12$, the variance of the old surmise intersects with the actual variance of the $k$NN spacing distribution, and thus the old and the corrected surmise give an equally good fit. For long-range spacings $k=17, \, 24$, we find that the old surmise predicts a narrower distribution---due to its smaller variance--- while the corrected surmise accurately captures the numerical data. 
Note that since the Skewness and excess Kurtosis of these distributions is small---cf. Fig. \ref{fig:skew_and_kurt}---the distributions look Gaussian, and indeed, the Gaussian distribution Eq.~\eqref{Pk_Gauss} also provides a good model for large $k\gtrsim 10$.  \change{Finally, we checked that using the expression Eq.~\eqref{alpha_new_long} for $\tilde{\alpha}$ instead of its large $k$ approximation, Eq.~\eqref{alpha_new}, improves the fit slightly for a few small value of $k$ as follows: for GOE it provides a better fit for $k=2,3$ with $\sigma=0.29, 0.37$ instead of $0.36, 0.38$, respectively; for GUE it provides a better fit for $k=2$ with $\sigma=0.38$ instead of $0.41$; and for GSE we found no improvement. For higher values of $k$ we found the same value of $\sigma$ when rounding to two decimal places.}

To summarize, we find that for all three Gaussian ensembles for $k\geq 2$, the new surmise always performs better \change{than the old surmise}.

\section{The XXZ model with random disorder}
\label{sec:XXZ_numerics}

In this Section, we test our new surmise for the $k$NN spacing distributions on a physical model which exhibits a transition from chaos to integrability. We show how the deviation from \pablo{the corrected surmise} witnesses the breakdown of \pablo{random matrix} universality. The model we use is the Heisenberg XXZ spin chain 
\begin{eqnarray}
    H_\textsc{xxz} = \sum_{n=1}^L ( S_n^x S_{n+1}^{x}+ S_n^y S_{n+1}^{y} +J_z  S_n^z  S_{n+1}^{z} ), ~~~~~
\end{eqnarray}
to which random magnetic fields are added on each site
\begin{eqnarray} \label{H_dis}
    H_{\text{dis}} = \sum_{n=1}^L h_n^z S_n^z.
\end{eqnarray}
Here $h_n^z$ are real random numbers taken from a uniform distribution, $\mathcal{U}_{[-W/2,W/2]}$, of width $W$.

The Hamiltonian $H \equiv H_\textsc{xxz}+H_{\text{dis}}$ conserves the total spin in the $z$-direction, $\mathcal{S}^z = \sum_{n=1}^L S_n^z$. 
We choose to work in the sector with half of the spins up and half of the spins down, which is of dimension ${{L}\choose{L/2}}$.  We present results for an open spin chain with $L=16$ spins. We draw our statistics from 100 realizations of disorder for $W=1,20$ and 50 realizations for other values of the disorder strength, $W$.

As explained in App.~\ref{App:unfolding}, for each disorder realization, we select eigenvalues \change{in} a spectral window \change{from the densest part of the spectrum} for which the density of states is not less than 0.9 times the maximal density\change{; this ensures that the window we choose is in the bulk of the spectrum and away from the spectrum's edges. We then} perform the unfolding procedure in that window. \change{The number of energy eigenvalues selected in this way is of the order of a few thousands out of the total number of $12860$ eigenvalues for each realization (the exact number depends on the value of $W$ and on the particular disorder realization). }

\begin{figure*}
    \centering
    \includegraphics[width=1\linewidth]{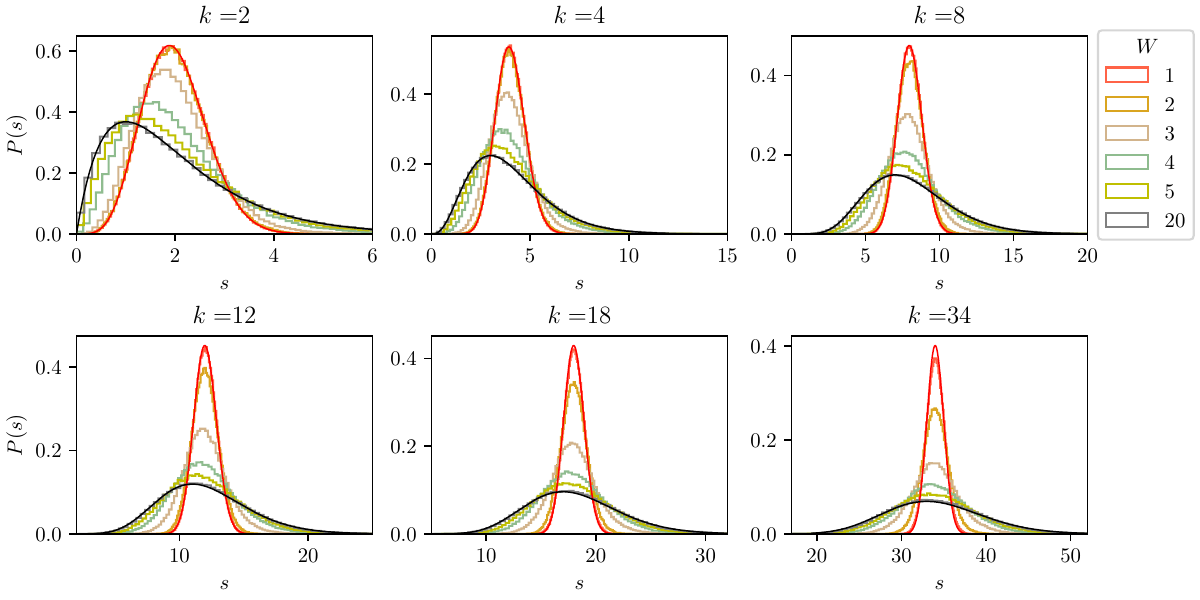}
    \caption{Numerical data for the $k$NN distributions for the XXZ spin chain with random on-site magnetic fields (step histograms) For \change{$W=1,2,3,4,5$ and $20$}. The red smooth curve shows the corrected surmise Eq.~\eqref{Pk_new} and the black smooth curve shows the $k$NN distributions for the 1d Poisson point process Eq.~\eqref{Pk_Poisson}.  We see a refined picture of how the spectral statistics change from chaos to integrability at various spectral ranges. }
    \label{fig:XXZ_kNN}
\end{figure*}

This model is often used in the context of many-body localization, see e.g. \cite{vznidarivc2008many, pal2010many, serbyn2016spectral, bertrand2016anomalous, sierant2019level, sierant2020model}, and is known to transition from chaotic to integrable spectral behavior as the disorder strength $W$ is increased. The Hamiltonian is real and symmetric and thus, in the chaotic regime, we expect the spectral $k$NN distributions of this model to coincide with those of a random matrix from the GOE. 

Figure \ref{fig:XXZ_variance} (left) shows the goodness of fit $\sigma$ between the $k$NN distribution for GOE and the chaotic phase $W=1, 1.25, 1.5, 1.75, 2$ of the disordered XXZ model. We see that for $W=1$ the $k$NN distribution of the GOE provides a very good model for $k\lesssim 15$\aurelia{, beyond which range the standard deviation starts to grow}. This indicates the breakdown of universality for long-range correlations. When the disorder $W$ increases, the value of the standard deviation $\sigma$ stays constant for small values of $k$ and then  deviates for larger values of $k$. \aurelia{The more we deviate from the chaotic phase, the smaller the value of $k$ at which the transition between these two regimes happens.} The slope of the $\sigma$ curve also increases with $W$. Notably, the breakdown of universality happens \aurelia{with very small increase of the disorder}. Indeed, for $W=2$ already, the only spectral ranges that resemble GOE are $k=2,3$.

Figure \ref{fig:XXZ_variance} (middle) shows the variance of the $k$NN distributions for various values of the disorder strength $W$. For reference, we plot the variance values expected from random matrices from the GOE (red) as well as the variances expected from the $k$NN distribution of the 1d Poisson point process (black). The plot shows that the smallest values of the disorder $W = 1, \, 1.25, \, 1.5, 1.75 $ follow the predictions of GOE until a certain $k$, after which the variance starts to grow faster than GOE. This \pablo{also} showcases the breakdown of universality for long-range correlations. The critical spacing range $k$ at which this breakdown occurs progressively decreases with the strength of the disorder $W$. For completeness, we also plotted the number variance along with the discrete values of $\Delta^{(k)}$. The difference between the number variance and the variance of the $k$NN distribution is shown in Figure \ref{fig:XXZ_variance} (right). For $W=1, 1.25, 1.5, 1.75$, this difference is always close to $1/6$. This indicates that the variance of the $k$NN distribution and the number variance always differ by a constant, but both deviate from the universal RMT prediction at the same rate \aurelia{for all ranges. In turn, 
 in the transition between chaos and integrability (intermediate values of $W$),} the difference between these quantities grows with $k$, indicating that the transition between chaos and integrability is monitored differently by these two quantities. Indeed, both quantities grow, but the number variance grows slightly faster for large $k$. 
\pablo{Lastly, in the integrable phase of the model, $W=20$, the difference between the two quantities is close to zero, as expected from the Poisson ensemble. 
}
\ruth{The fluctuations however are larger than in the chaotic regime possibly because the values of both quantities are much larger. }

\begin{figure}
    \centering
    \includegraphics[width=0.9\linewidth]{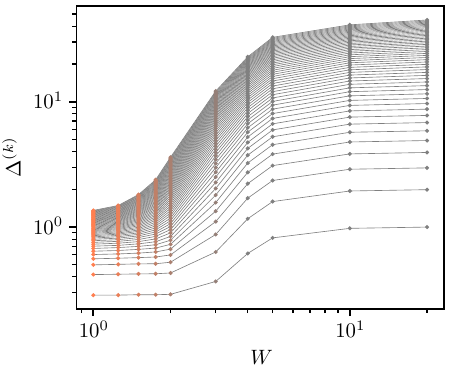}
    \caption{ \change{Numerical data for the variance $\Delta^{(k)}$ of the $k$NN distributions as a function of the disorder strength, $W$, for the XXZ spin chain with random on-site magnetic fields. The different lines show different $1\leq k \leq 50$ from bottom to top.}}
    \label{fig:Var_vs_W}
\end{figure}

Figure \ref{fig:XXZ_kNN} presents the full $k$NN distributions for several values of $k$ and $W$. As expected, at $W=1$, the model is in the chaotic regime with the distributions close to those of GOE. Deviations from the GOE distributions can be seen as $k$ becomes larger; this can also be seen in the plot of the variance, Fig. \ref{fig:XXZ_variance}. At $W=20$, the model is in the integrable regime and follows the Poisson $k$NN distributions with deviations at large $k$, as can be seen also in the plot of the variance, Figure \ref{fig:XXZ_variance}. 
At values in between these two values of disorder strength, we see spectral behavior between the two regimes for all values of $k$. \change{Finally, Figure~\ref{fig:Var_vs_W} shows how the variance $\Delta^{(k)}$ changes as a function of $W$ for $1\leq k \leq 50$. The crossover from chaos to integrability seems to occur at smaller $W$ for larger $k$, indicating a faster loss of correlation between further away energy eigenvalues. This shows that different neighbor ranges transition from chaos to integrability at different disorder strengths $W$. It would be interesting to test this result for larger systems and see whether such a $k$-dependent transition persists. In any case, this highlights the importance of studying longer-range correlations when looking at the transition/crossover from chaos to integrability.} 

\section{Summary}
\label{sec:summary}

The $k$NN distribution for Gaussian Random Matrices can be approximated by an expression similar to the Wigner surmise. In this work, we have shown how the  Wigner-like surmise suggested in previous literature fails to accurately capture both short- and long-range spectral correlations, as can be seen by inspecting its variance in Figure \ref{fig:variance_and_mean}. 
We provided a new surmise for the distribution of $k$NN spacings of random matrices by correcting the power \ruth{to a new power} $\tilde \alpha$ \ruth{given by Eq.~\eqref{alpha_new}}. The expression now reproduces the value of the variance of the $k$NN distributions for random matrices, and fits numerical random matrix data, as shown quantitatively using the standard deviation goodness-of-fit $\sigma$.  
We find 
that the corrected surmise consistently provides a good fit to the numerical $k$NN distributions already for $k\geq 2$. 
\change{We also found that the full expression, Eq.~\eqref{alpha_new_long} for $\tilde{\alpha}$, provides a slightly better fit for $k=2,3$ in the GOE and for $k=2$ in the GUE.}

\change{The number variance and the variances of the $k$NN distributions provide us with a single $k$-dependent number to study long-range correlations among eigenvalues. 
The full distributions give us more and allow us to study in more detail the structure of the correlations between eigenvalues and the breakdown of universality for long-range spectral correlations in physical systems.}

These results can be used as a refined probe of quantum chaos for testing longer-range correlations among eigenvalues, as exemplified by our numerical results for the XXZ spin chain with on-site disorder. We were able to study the full $P^{(k)}(s)$ distributions at various values of disorder strength, $W$.
The variance of these distributions can be used instead of the number variance, and, from a practical point of view, may be easier to compute numerically. \change{As Figure~\ref{fig:Var_vs_W} suggests, the variance of the $k$NN distributions can be used to quantitatively study the crossover from chaos to integrability as a function of the spectral distance $k$. It would be interesting to study the scaling of this $k$-dependent transition as a function of system size.}

In light of our results, the relationship between the number variance and the variance of the $P^{(k)}(s)$ distributions remains an open problem \cite{RevModPhys.53.385}. It would also be interesting to extend the $k$NN distributions to study  dissipative quantum chaos, where the complex spectrum hinders some long-range signatures of quantum chaos~\cite{garciagarcia_universality_2023, xu_chaosvsdecoh_2021, roccati_NHMBL_2024, akemann_transitions_2025}. 

\begin{acknowledgements}
We thank \aurelia{Gernot Akemann and Patricia P{\"a}{\ss}ler} for insightful discussions and acknowledge funding from the Luxembourg National Research Fund (FNR, Attract grant QOMPET 15382998). The numerical simulations presented in this work were carried out using the HPC facilities of the University of Luxembourg.
\end{acknowledgements}

\onecolumngrid
\appendix

\counterwithin{figure}{section}

\section{Surmise derivation} 
\label{App:surmise_derivation}

To find a surmise for the distribution of the $k$th neighbor level spacing, we use a $N=k+1$ dimensional random matrix with energy levels $E_1 \leq E_2 \leq \dots \leq E_{k+1}$ and attempt to compute 
\begin{align}\label{eq:Pks-def}
    P^{(k)}(s) &= \int_{-\infty}^\infty \!\! dE_1 \int_{E_1}^\infty \!\! dE_2 \, \dots  \int_{E_k}^\infty \!\! dE_{k+1}  \rho(E_1,\dots,E_{k+1})      \delta\big[s-(E_{k+1}-E_1)\big]. 
\end{align}
The integration limits take into account the ordering of the levels and $\rho$ is the joint eigenvalue probability distribution
\begin{equation} \label{JPD_eigenvalues}
  \!\!  \rho(E_1,\dots, E_N)= C \!\!\!\!\!\!\! \prod_{1\leq i <j\leq N} \!\!\!\!\!\! |E_i-E_j|^\beta \, e^{{-}A\!\sum_{i=1}^N \!\! E_i^2}.\!\!\!
\end{equation}
Following the derivation in \cite{kahn_statistical_1963}, we change variables from $\{E_1,E_2,\dots,E_{k+1}\}$ to $\{E_1, s_1, s_2, \dots, s_k\}$, where $s_i\equiv s_i^{(1)}$ are the NN spacings. Then
\begin{align}
    P^{(k)}(s) &\propto \int_{-\infty}^\infty dE_1 \int_0^\infty ds_1 \dots  \int_0^\infty ds_k   \rho(E_1,s_1\dots,s_k) \delta\Big(s-\sum_{i=1}^k s_i\Big).
\end{align}
Note that $E_1$ appears only in the exponential term of $\rho(E_1,s_1\dots,s_k)= \big[p(s_1,s_2,\dots, s_k)\big]^\beta     \times e^{-A[ E_1^2 +(E_1+s_1)^2+ (E_1+s_1+s_2)^2+\dots + (E_1+s_1+\dots+s_k)^2]}$, 
written here using the $k(k+1)/2$ degree polynomial: 
\begin{align} \label{poly}
    p(s_1,s_2,\dots, s_k) &{=} s_1(s_1{+}s_2)\dots (s_1{+}s_2{+}\dots{+}s_k)
    \nonumber\\
    &{\times} s_2 (s_2{+}s_3) \dots (s_2{+}s_3{+}\dots{+}s_k)
    {\times} \dots {\times} s_{k-1}(s_{k-1}{+}s_k) {\times} s_k ~.\!\!\!
\end{align}
Since $s_i\geq 0$ for all $i$, this polynomial is positive everywhere.
Performing the Gaussian integral over $E_1$, we find (up to a constant):
\begin{align}
    P^{(k)}(s) &\propto  \int_0^\infty ds_1 \dots  \int_0^\infty ds_k  \big[p(s_1,s_2,\dots, s_k)\big]^\beta
    e^{-A\,q(s_1,\dots,s_k)}\,\delta\Big(s-\sum_{i=1}^k s_i\Big), 
\end{align}
where we defined the quadratic polynomial resulting from the integration over $E_1$ as 
\begin{equation} \label{quad}
    q(\{s_i\}) = \! \sum_{i=1}^{k} \frac{i(k{+}1{-}i)}{k{+}1} s_i^2  +\!\!\!\sum_{i<j=1}^{k} \frac{2i(k{+}1{-}j)}{k{+}1}s_is_j~.
\end{equation} 
Note that since $s_i\geq 0$ for all $i$ and the coefficients of this quadratic polynomial are always positive, $q(s_1,\dots,s_k)\geq 0$ everywhere. 
Next, we rescale the spacings as $x_i = s_i/s$. Taking into account the Jacobian of this transformation, which is $s^k$, the homogeneity of $p(s_1,\dots,s_k)$ and of $q(s_1,\dots, s_k)$, and using the delta function identity $\delta(a x) = \delta(x)/|a|$, we arrive at
\begin{align}
    P^{(k)}(s) &\propto  s^{k-1} s^{\frac{k(k+1)}{2}\beta}\int_0^\infty dx_1 \dots  \int_0^\infty dx_k  
    \big[p(x_1,x_2,\dots, x_k)\big]^\beta e^{-A s^2 q(x_1,\dots,x_k)}
    \delta\Big(1{-}\!\!\sum_{i=1}^k x_i\Big).
\end{align}
This is an integral over a \textit{$(k-1)$ simplex}.  Using the $\delta$-function to set $x_k=1-x_1-x_2-\dots-x_{k-1}$ and restricting the integration limits, we replace the quadratic function by $\overline{q}(\{x_i\}_{i=1}^{k-1})=\frac{k}{k+1}-2\sum_{i=1}^{k-1}B_ix_i+\sum_{i,j=1}^{k-1}A_{ij}x_ix_j$ where
the elements $B_i$ and $A_{ij}$ for $i,j=1,\dots,k-1$ are given by \footnote{It can be checked that $\mathbf{A}$ is positive definite. The inverse of $\mathbf{A}$ is a tridiagonal, almost Toeplitz matrix, with diagonal $A^{-1}_{11}=3/2$ and $A^{-1}_{ii}=2$ for $i\neq 1$, and off-diagonals $A^{-1}_{i,i+1}=A^{-1}_{i-1,i}=-1$. It can also be verified that $\mathbf{m}=\mathbf{A}^{-1}\mathbf{B}=(1/2,0,0,\dots,0)$.}
\begin{subequations}
\begin{align}
    B_i &= \frac{k-i}{k+1}, \\
    A_{ii} &= \frac{(k-i)(i+1)}{k+1}, \\
    A_{i\neq j} &= \frac{(k-\max(i,j))(\min(i,j)+1)}{k+1}~.
\end{align}
\end{subequations}
Similarly, the polynomial $p(\{x_i\}_{i=1}^{k})$ is replaced by $\overline{p}(\{x_i\}_{i=1}^{k-1})$.  After completing the square, we arrive at the $(k-1)$-dimensional integral 
\begin{align}\label{exact_Pk_s_simplex}
    &P^{(k)}(s) \propto s^{\frac{k(k+1)}{2}\beta}s^{k-1} e^{-\frac{A}{2} s^2} 
    \int_0^1 dx_1 \int_0^{1-x_1}dx_2 \dots  \int_0^{1-\sum_{i=1}^{k-2}x_i} dx_{k-1}   
    \overline{p}(\mathbf x)^\beta e^{- A s^2(\mathbf{x}^T-\mathbf{m}^T)\mathbf{A}(\mathbf{x}-\mathbf{m})}, 
\end{align}
where we introduced the vectors $\mathbf{m}=\mathbf{A}^{-1}\mathbf{B}$, $\mathbf x = (x_1, \dots, x_{k-1})$ and where we have used $A\big[\frac{k}{k+1}- \mathbf{B}^T\mathbf{A}^{-1}\mathbf{B}\big]=A/2$ for all $k$.  The vector $\mathbf{m}$ simplifies to $(1/2,0,\dots,0)$ which means that the Gaussian is centered at zero for all $x_i$ except $x_1$. 

The integral over the simplex is challenging to compute exactly. For small $s$, the Gaussian in the integral can be expanded to second order in $s$, resulting in a correction to the width of the pre-factor Gaussian function, while at large $s$ the integral gives corrections to the power law.
\aurelia{
We thus end up with the generalized distribution given in Eq.~\eqref{Pk}.
Since similar results have been reported in the literature, let us briefly review their arguments here. 
The first generalization of Wigner's surmise, that we are aware of, assumes a Brody-like ansatz, which leaves the power-law $\alpha$ as a free parameter \cite{engel_higher-order_1998}. In Ref. \cite{abul-magd_high-order_2000}, the power-law in Eq.~\eqref{alpha_old} is found using a small $s$ expansion and the generalized Wigner surmise, Eq.~\eqref{Pk}, is obtained assuming a Gaussian behavior at large $s$. This approach is also followed in \cite{sakhr_wigner_2006} in the context of 2-dimensional Poisson point processes. These references thus find the same distribution through heuristic arguments. 
Formal results for the $k$NN probability distributions can be obtained exactly using tools from RMT, see e.g. \cite{mehta2004random}, and there are even connections between the different $k$NN distributions \cite{forrester_random_2009}. However, these formal and exact results lack an explicit expression for the $k$NN distribution reminiscent of the Wigner surmise, which is itself an approximation \cite{Haake:2010fgh}.
More recently, an extension of these results to spacing ratios was tested numerically \cite{tekur_higher-order_2018} but with no analytical proof. Lastly, Rao \cite{rao_higher-order_2020} proposed an analytical derivation of the generalized Wigner distribution from the joint-probability-density of eigenvalues. However, since the energies are not ordered, the spacing $E_{k+1}-E_1$ need not be a $k$-th level spacing. The same work also seems to suggest that the generalized Wigner distribution is the exact distribution of the $k$NN spacing, despite the latter being known as an approximation only, as can already be seen in exact small size results like \cite{kahn_statistical_1963, berry_geometric_2018}. In turn, our derivation is based on Wigner's original argument, i.e. considering the joint-probability-density of eigenvalues of the largest possible matrix with a $k$-th level spacing; we have explicitly stated the approximations involved in obtaining the generalized Wigner distribution and discussed the corrections to the distribution.
}

\section{Unfolding} \label{App:unfolding}

The unfolding procedure ensures that only local fluctuations are taken into account when computing spectral correlations among eigenvalues. This is achieved by removing the energy density dependence on the global density of states, $\bar{\rho}(E)$, which is system dependent. 

As described in \cite{Haake:2010fgh, Guhr:1997ve}, the cumulative distribution function of the average density of states, $\bar{\rho}(E)$, multiplied by the number of eigenvalues in the desired energy window, 
\begin{eqnarray} \label{fE}
    f(E) = N \int_{-\infty}^E dE' \, \bar{\rho}(E')~,
\end{eqnarray}
provides the function that achieves a uniform average density of states. 

In the case of the Gaussian ensembles---GOE, GUE and GSE---the global density of states in the large $N$ limit is given by Wigner's semicircle \cite{mehta2004random} and the integral in Eq.~\eqref{fE} can be done analytically \footnote{\url{https://robertsweeneyblanco.github.io/Computational_Random_Matrix_Theory/Eigenvalues/Wigner_Semicircle_Law.html}}. In the case of a physical system such as the one we are considering in this work---the XXZ spin chain with random magnetic fields---the global density of states is unknown. Then, the integral is done numerically and fitted to a polynomial function for each realization of the disorder.  We focus on eigenvalues from the densest part of the spectrum by choosing those eigenvalues which lay within a region where the density of states is not less than $0.9$ of its maximal value. We then perform the integral in Eq.~\eqref{fE} for that set of eigenvalues and fit it to a third-degree polynomial.

\section{The 1d Poisson point process} \label{App:Pk_Pois}

The NN spectral statistics for uncorrelated energy eigenvalues is given by $P^{(1)}(s)=e^{-\bar{\rho}s}$, where $\bar{\rho}$ is the average density of states, taken to be the uniform distribution $\bar{\rho}(E)=\bar{\rho} = 1$. This is known as the `Poissonian' NN distribution. We will now present a derivation for the $k$NN distributions of the 1d Poisson point process (see also \cite{Sakhr_2006}).
Since there are no correlations between the energies, the joint probability distribution of the NNs, $\{s_i\}_{i=1}^k$, is a product of the individual NN distributions, 
$P(s_1,s_2,\dots, s_k) = P(s_1)P(s_2)\dots P(s_k) =e^{{-}\! \sum_{i=1}^k s_i} $.
Using the joint distribution, $P(s_1,s_2,\dots, s_k)$, we can define the probability that the $k$th nearest neighbor spacing is $s$ as 
\begin{equation}
   P^{(k)}(s) \propto \int_0^\infty \!\! ds_1  \dots ds_{k}\, P(\{s_i\}_{i=1}^k) \delta(s-\sum_{i=1}^k s_i) .\quad
\end{equation}
As in App.~\ref{App:surmise_derivation}, we rescale the variables, $x_i=s_i/s$. Taking into account the Jacobian, given by $s^k$, and the integrating over the delta function, which contributes a $s^{-1}$ and sets everywhere $x_k=1-\sum_{i=1}^{k-1}x_i$, we find:
\begin{eqnarray}
   P^{(k)}(s) &{\propto}& s^{k-1} e^{-s}\! \int_0^1 \!\! dx_1 \int_0^{1-x_1} \!\! dx_2 \dots \!\!\int_0^{1-\!\sum_{i{=}1}^{k{-}2}x_i}\!\!dx_{k-1} \nonumber\\
   &=& s^{k-1} e^{-s} \times \textrm{constant}~,
\end{eqnarray}
 where the second equality follows from the fact that the integral over the simplex is a finite constant.  The normalized distribution is then given by
 \begin{eqnarray}\label{Pk_Poisson}
    P^{(k)}_{\text{Poisson}}(s) &=& \frac{1}{(k-1)!}\, s^{k-1} e^{-s}~.
 \end{eqnarray} 
 It can be checked that $\langle s\rangle=k$, as expected, and that the variance of this distribution is $\Delta^{(k)} = k$.

\begin{figure}[]
    \centering    
    \includegraphics[width=0.3\linewidth]{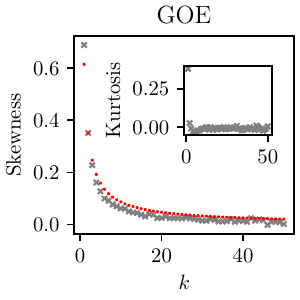}\hfill\includegraphics[width=0.3\linewidth]{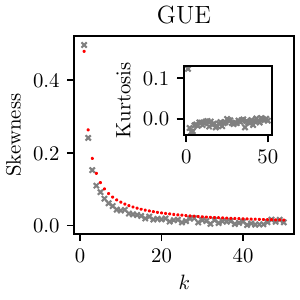}\hfill\includegraphics[width=0.3\linewidth]{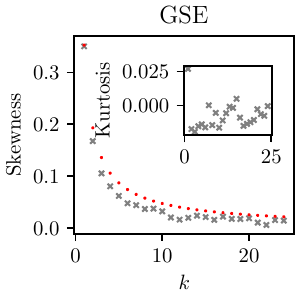}
    \caption{Skewness and excess Kurtosis (inset) of the $k$NN distributions for $k\geq 1$. Both tend to zero as $k$ is increased, implying that the distributions become Gaussian at large $k$. The analytical values of the Skewness (red dots) are computed using Eq.~\eqref{alpha_new} in Eq.~\eqref{Skew_alpha}. }
    \label{fig:skew_and_kurt}
\end{figure}

\section{Computational cost of computing the number variance vs the variance of the $k$NN distributions}
\label{App:NV_vs_variance}
\change{For a fixed value of $L$, the computation of the number variance requires going over the spectrum $N_{\xi_s}$ times, checking if a given $E_n$ is in the interval $[\xi_s, \xi_s + L]$ and computing the variance of $\eta(\xi_s, L)$. This procedure has to be repeated $N_{\rm av}$ times to average over the RMT ensemble or the disorder realizations. This means that we need to perform $O(N_{\rm av} N_{\xi_s} N)$ individual operations and we need to compute a variance $O(N_{\rm av})$ times. Comparatively, the $k$NN variance provides a better scaling, for a fixed $k$ we need to run over the spectrum $N_{\rm av}$ times computing the difference $E_{n+k} - E_n$ which gives $O(N_{\rm av} N)$ operations, and we only need to compute the variance a single time. Furthermore, the individual operations needed to compute the number variance involve two $\texttt{if}$ statements to check whether $E_n \geq \xi_s$ and $E_n \leq \xi_s + L$, while for the $k$NN variance the individual operation is a simple subtraction. Let us denote the complexity of each operation as: $\mathcal{C}_{\texttt{window}}$ to check if the energy is in the window $E_n \in [\xi_s, \xi_s + L]$, $\mathcal{C}_{\texttt{s}}$ for the subtraction and $\mathcal{C}_{\texttt{var}}(n)$ for the variance of $n$ elements. For a single $L$, the complexity of the number variance is then of the order
\begin{equation}
    \mathcal{C}_{\texttt{NV}} \sim   \mathcal{C}_{\texttt{window}}N_{\rm av} N_{\xi_s} N  + \mathcal{C}_{\texttt{var}}(N N_{\xi_s}) N_{\rm av} ,
\end{equation}
while for a single $k$NN variance it is 
\begin{equation}
    \mathcal{C}_{\texttt{kNNvar}} \sim \mathcal{C}_{\texttt{s}} N_{\rm av} N + \mathcal{C}_{\texttt{var}}(N N_{\rm av}),
\end{equation}
which shows that the $k$NN variance, although related to the number variance, is less computationally costly to compute.
}

\section{Skewness and Kurtosis}
\label{App:Skew_Kurt}

To see whether the considered distributions are normal, we also look at the Skewness and excess Kurtosis---which are expected to be zero for a normal distribution. 
For the $k$NN distributions, they are  
\begin{equation}
    \mathrm{Skew}^{(k)} = \frac{\langle(s^{(k)} -\langle s^{(k)} \rangle)^3\rangle}{(\Delta^{(k)})^{3/2}}
\end{equation}
and
\begin{equation}
\mathrm{Kurt}^{(k)} = \frac{\langle(s^{(k)} -\langle s^{(k)} \rangle)^4\rangle}{(\Delta^{(k)})^2}-3,
\end{equation} 
respectively. For the distribution Eq.~\eqref{Pk}, the Skewness reads
\begin{eqnarray}
\mathrm{Skew}^{(k)}\!&\!=&\!\frac{\sqrt{2}\! \left(\!4 \Gamma\! \left(\frac{\alpha }{2}\!+\!1\right)^3\!-\!(2 \alpha \!+\!1) \Gamma\! \left(\frac{\alpha }{2}\!+\!1\right)\! \Gamma\!
   \left(\frac{\alpha\! +\!1}{2}\right)^2\!\right)}{\left((\alpha +1) \Gamma \left(\frac{\alpha +1}{2}\right)^2-2 \Gamma \left(\frac{\alpha
   }{2}+1\right)^2\right)^{3/2}}  \label{Skew_alpha}\\
   &=&\frac{1}{\sqrt{2 \alpha}}+  O(\alpha^{-3/2}).
\end{eqnarray}
See Figure \ref{fig:skew_and_kurt} for numerical data for GOE, GUE and GSE versus analytical data from Eq.~\eqref{Skew_alpha}, computed using Eq.~\eqref{alpha_new}.

It is also interesting to look at the Skewness of the $k$NN distributions from the 1d Poisson point process, Eq.~\eqref{Pk_Poisson}. It is given by $\mathrm{Skew}^{(k)} = \frac{2}{\sqrt{k}}$. Note that it goes to zero, as a function of $k$, much slower than the result for RMT, which goes as $1/\sqrt{2\alpha}$ indicating a $\sim 1/k$ decay for $\alpha$ given by Eq.~\eqref{alpha_old} and $\sim \sqrt{\ln k}/k$ for Eq.~\eqref{alpha_new}.

\section{A better surmise for the $k=1$ case}\label{App:Pk1_WS}

\begin{figure}[h]
    \centering
    \includegraphics[width=0.7\linewidth]{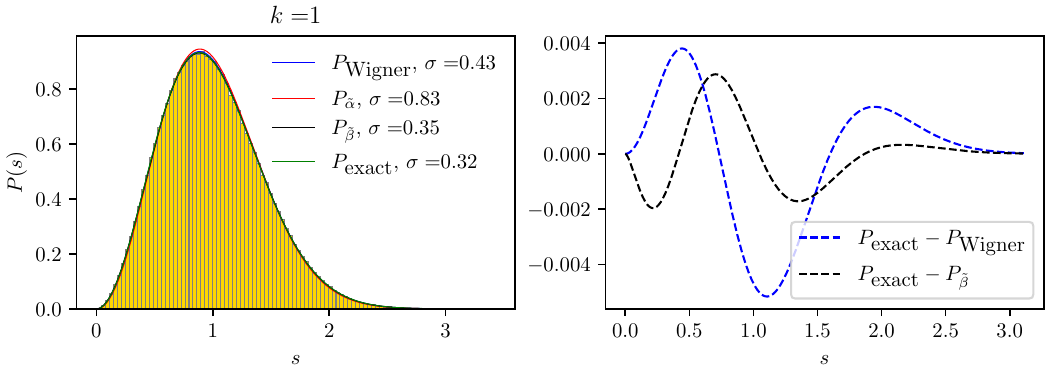}
    \caption{ \change{(Right) numerical data from 1000 realizations of GUE matrices of size $N=1000$ (histogram) is compared with:  Wigner's surmise (blue line), the corrected surmise $P_{\tilde{\alpha}}$ given by Eq.~\eqref{Pk_new} at $k=1$ (red line), $P_{\tilde \beta}$ given by Eq.~\eqref{NN_WS} with corrected power Eq.~\eqref{beta_new} (black line). We also show the exact result from \cite{dietz1990taylor} (green line). (Left) The difference between the exact result and Wigner's surmise (dashed blue line), and the difference between the exact result and Eq.~\eqref{NN_WS} with corrected power Eq.~\eqref{beta_new} (dashed black line). Both the goodness-of-fit, $\sigma$, and the difference plot indicate that using the corrected power $\tilde{\beta}$ in Eq.~\eqref{NN_WS} performs better than Wigner's surmise (given by Eq.~\eqref{NN_WS} with $\beta=2$). }}
    \label{fig:Pk1}
\end{figure}

\change{In this Appendix we discuss the NN level spacing distribution, focusing on the GUE. 
The corrected power, $\tilde\alpha$, was derived from Eq.~\eqref{Eq_for_talpha} using the large $\alpha$ expansion of $A_\alpha$, Eq.~\eqref{A_approx}.  Another source of inaccuracy in $\tilde\alpha$ comes from the initial condition which is the variance as derived from Wigner's surmise -- which is in itself inexact. It is thus not expected to work particularly well at $k=1$.
In Figure \ref{fig:Pk1} we compare the performance of the new surmise, Eq.~\eqref{Pk_new} with $\tilde{\alpha}$ given by Eq.~\eqref{alpha_new} \footnote{\change{Incidentally, at $k=1$ Eq.~\eqref{alpha_new} for $\tilde\alpha$ performs better than the long expression Eq.~\eqref{alpha_new_long}. This can happen since $k=1$ is away from large $k$ for which $\tilde \alpha$ was derived.}}, with that of Wigner's surmise and with the exact result at large dimension given in \cite{dietz1990taylor}. The exact result is a product of Dyson's asymptotic result at large $s$ and a Pad\'{e} expansion evaluated from a small $s$ Taylor expansion of the exact level spacing distribution. Dyson's asymptotic result is given by:
\begin{eqnarray}
    P_\text{as}(s) = \frac{\pi^2}{16}\left(s^2-\frac{2}{\pi^2}+\frac{5}{\pi^2s^2} \right)E_\text{as}(s)
\end{eqnarray}
where $E_\text{as}(s) = (\pi/2)^{-1/4}e^{2B}s^{-1/4}e^{-\pi^2 s^2/8}$ with $B = \ln(2)/24+ (3/2)\zeta'(-1)$ \footnote{Here, $\zeta'(z)$ is the derivative of the Riemann zeta function.}. The Pad\'{e} expansion is given by:
\begin{eqnarray}
    \text{Pad\'{e}}(s) = \frac{\sum_{m=0}^{2L} \nu_m s^{m/4}}{\sum_{m=0}^{2L}\Delta_m s^{m/4}}
\end{eqnarray}
where the coefficients $\nu_m$ and $\Delta_m$ can be computed from the first $L$ Taylor coefficients, and are given in Table 2 of \cite{dietz1990taylor} for $L=22$. Finally, the exact result can be expressed as:
\begin{eqnarray} \label{Ps_exact_GUE}
    P_\text{exact}(s) = \text{Pad\'{e}}(s)\,P_\text{as}(s).
\end{eqnarray}

Using the exact result, we can exploit the relation, Eq.~\eqref{variance}, between the variance and the power $\alpha$ without resorting to the large $k$ approximation to obtain an improvement over Wigner's surmise. This can be done by solving Eq.~\eqref{variance} numerically for $\alpha$ using the variance of the exact NN distribution as given by Eq.~\eqref{Ps_exact_GUE}, $\Delta^{(1)}_\text{exact} = 0.17999$. This computation gives a corrected power 
\begin{eqnarray} \label{beta_new}
    \tilde \beta=1.96998
\end{eqnarray}
to use in Eq.~\eqref{NN_WS} instead of $\beta=2$.

Figure \ref{fig:Pk1} (left) shows Wigner's surmise, Eq.~\eqref{NN_WS} with $\beta=2$, the corrected surmise, Eq.~\eqref{Pk_new} at $k=1$, and Eq.~\eqref{NN_WS} with the corrected power given by Eq.~\eqref{beta_new}.
We also show the exact analytical result in terms of a Pad\'{e} expansion, Eq.~\eqref{Ps_exact_GUE}. We compute the $\sigma$ goodness-of-fit for each of these results compared with numerical data in a range between $0$ and $4$ standard deviations from the mean of the distribution (which is unity in this case).
The goodness-of-fit using $\tilde{\beta}$ in Eq.~\eqref{NN_WS} indeed performs better than Wigner's surmise (with $\beta=2$). In the right panel of this figure we compare the differences between: the exact result and Wigner's surmise (reproducing Figure 2 in \cite{dietz1990taylor}), and the exact result and Eq.~\eqref{NN_WS} with $\tilde \beta$ given by Eq.~\eqref{beta_new}.}

\section{Results for the GOE and GSE}
\label{App:GOE_GSE}

\begin{figure}
    \centering \includegraphics[width=1\linewidth]{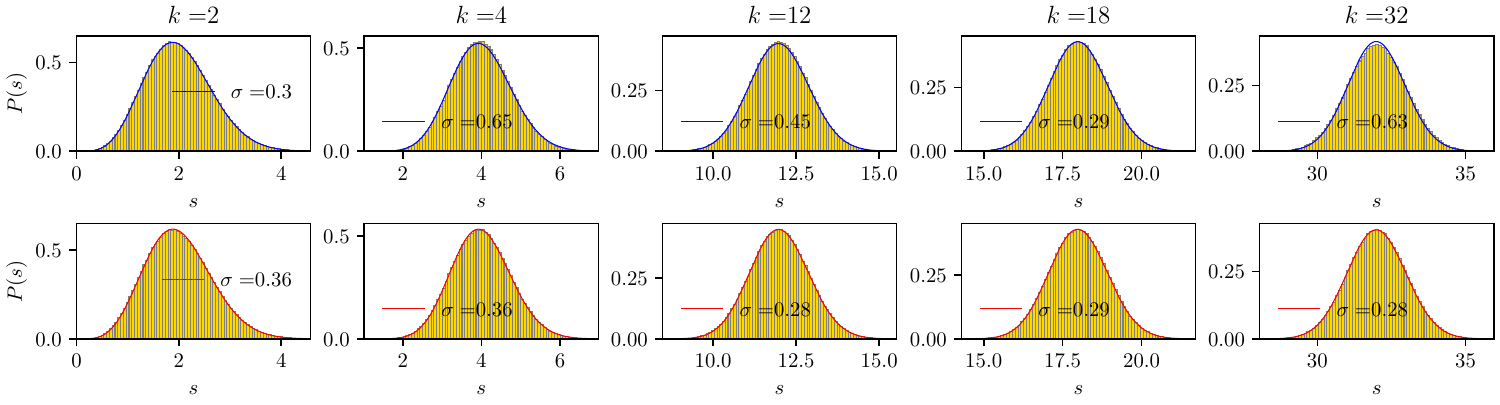}
    \caption{The $k$NN distributions for a few values of $k$, for the GOE. The histograms show the numerical data for 1000 realizations of size $N=1000$ matrices. The top row shows the old surmise (blue curves) and the bottom row shows the new surmise (red curves).  \ruth{The standard deviation goodness-of-fit, $\sigma$, between the analytical distributions and the numerical data is shown in units of $10^{-2}$.}}
    \label{fig:Pk_GOE}
\end{figure}

\begin{figure}
    \centering
     \includegraphics[width=1\linewidth]{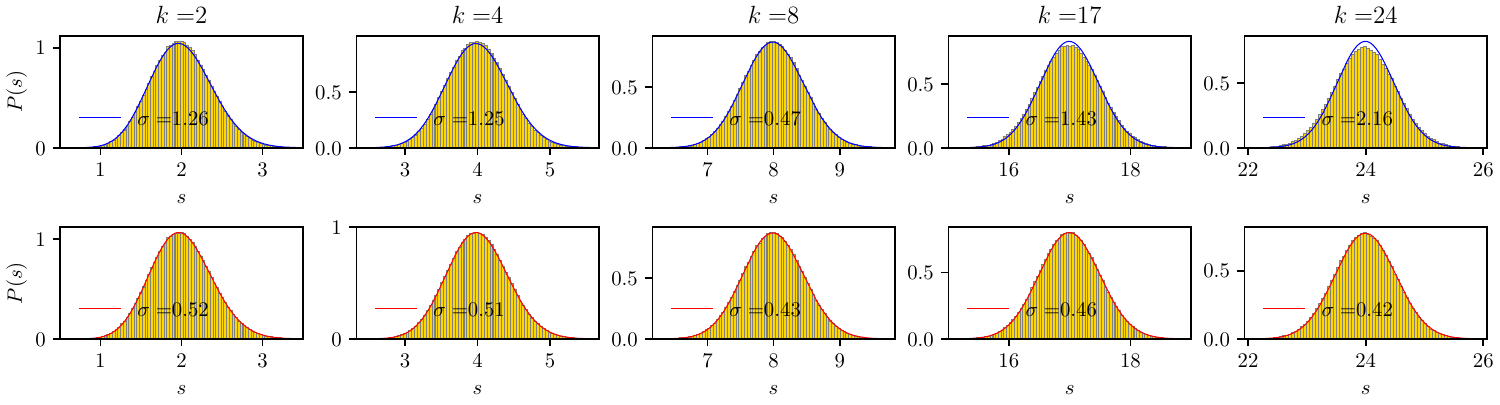}
    \caption{The $k$NN distributions for a few values of $k$, for the GSE. The histograms show the numerical data for 1000 realizations of size $N=2000$ matrices. The top row shows the old surmise (blue curves) and the bottom row shows the new surmise (red curves). \ruth{The standard deviation goodness-of-fit, $\sigma$, between the analytical distributions and the numerical data is shown in units of $10^{-2}$.}}
    \label{fig:Pk_GSE}
\end{figure}

In this Appendix, we show the explicit $P^{(k)}(s)$ distributions for some values of $k$ for the GOE and the GSE. Figures \ref{fig:Pk_GOE} and \ref{fig:Pk_GSE} compare the old surmise and the new surmise with numerical results for these ensembles. For the complete set of $\sigma$ goodness-of-fit values, see Figure \ref{fig:SD} in the main text. \pablo{Similarly to what we observed for the GUE in the main text, the corrected distributions for the GOE and GSE consistently perform better than the old surmise; as discussed in the main text, using $\tilde{\alpha}$ given by Eq.~\eqref{alpha_new_long} further improves the fit for the GOE at $k=2,3$. Below a certain value, $k=18$ in GOE and $k=8$ in GSE, the old surmise is slightly smaller at the peak, due to its slightly bigger variance than the actual RMT result. After this point the old surmise overestimates the value at the peak, due to its smaller variance. At the crossing point, \ruth{$k=18$ for GOE and $k=8$ for GSE}, the old and the corrected surmises perform equally well in describing the numerical distribution.}

%
\twocolumngrid

\bibliographystyle{apsrev4-1}
\bibliography{references}

\end{document}